\documentclass[a4paper,11pt]{article}
\usepackage{latexsym}
\usepackage{amsmath, amsthm, amssymb, bbm}
\usepackage[mathscr]{eucal}
\usepackage{cite}
\usepackage{hyperref}
\usepackage[margin=2cm]{geometry}
\usepackage{authblk}

\theoremstyle{plain}

\theoremstyle{definition}

\theoremstyle{remark}

\newcommand{\lhs}{l.h.s.\ }
\newcommand{\wrt}{w.r.t.\ }
\newcommand{\cf}{cf.\ }

\newcommand{\bra}[1]{\langle #1 \rvert}
\newcommand{\ket}[1]{\lvert #1 \rangle}

\newcommand{\VEV}[1]{\langle #1 \rangle}

\newcommand{\ud}{\mathrm{d}}
\newcommand{\del}{\partial}

\DeclareMathOperator{\Tr}{Tr}
\DeclareMathOperator{\diag}{diag}

\newcommand{\R}{\mathbb{R}}

\newcommand{\Z}{\mathbb{Z}}

\newcommand{\order}{\mathcal{O}}

\newcommand{\1}{\mathbbm{1}}

\newcommand{\sS}{\mathcal{S}}

\newcommand{\cR}{\mathcal{R}}

\newcommand{\cM}{\mathcal{M}}
\newcommand{\cN}{\mathcal{N}}
\newcommand{\cB}{\mathcal{B}}
\newcommand{\MM}{\mathcal{M}}
\newcommand{\NN}{\mathcal{N}}
\newcommand{\BB}{\mathcal{B}}
\newcommand{\cA}{\mathcal{A}}

\newcommand{\eps}{\varepsilon}

\newcommand{\pd}{{\del}}
\newcommand{\ii}{i}

\newcommand{\N}{\mathbb{N}}

\newcommand{\vp}{{\varphi}}

\newcommand{\nn}{\nonumber}
\newcommand{\beq}{\begin{equation}}
\newcommand{\eeq}{\end{equation}}

\newcommand{\defeq}{\mathrel{:=}}

\begin{document}

\title{The semi-classical energy of closed Nambu-Goto strings}
\author[1,2]{Marek Kozo\v{n}\thanks{m.kozon@utwente.nl; Currently at Mathematics of Computational Science (MACS) \& Complex Photonic Systems (COPS), MESA+ Institute for Nanotechnology, University of Twente, Enschede, The Netherlands}}
\author[2]{Jochen Zahn\thanks{jochen.zahn@itp.uni-leipzig.de}}

\renewcommand\Affilfont{\itshape\small}
\affil[1]{Institute of Theoretical Physics, Faculty of Mathematics and Physics, Charles University in Prague, V~Hole{\v s}ovi{\v c}k{\' a}ch 2, CZ-180 00 Praha 8, Czech Republic}
\affil[2]{Institut~f\"ur~Theoretische~Physik,~Universit\"at~Leipzig,~Br\"uderstr.~16,~04103~Leipzig,~Germany}

\date{\today}

\maketitle

\begin{abstract}
We compute semi-classical corrections to the energy of rotating closed Nambu-Goto strings. We confirm the results obtained by means of the Polchinski-Strominger action. We also show that in this semi-classical approximation, the spectrum of physical excitations contains modes that are unphysical non-perturbatively, i.e., to which no physical excitations of the covariantly quantized Nambu-Goto string correspond.
\end{abstract}

\section{Introduction}

In the covariant quantization scheme, the Regge intercept $a$ of the Nambu-Goto string is a free parameter, only constrained by $a \leq 1$ for $D \leq 25$ and $a =1$ for $D = 26$, with $D$ the dimension of the target Minkowski space \cite{Rebbi74, Scherk75}. As the Nambu-Goto string is a toy model for vortex lines in QCD, \cf \cite{SonnenscheinWeissman14, SonnenscheinWeissman15, BrandtMeineri16} for recent works and reviews in this direction, a determination of $a$ for non-critical dimensions is desirable. Using the Polchinski-Strominger action \cite{PolchinskiStrominger}, Hellerman and Swanson \cite{HellermanSwanson} computed the Regge intercepts
\begin{align}
\label{eq:a_HS}
 a_{\mathrm{open}} & = 1, &
  a_{\mathrm{closed}} & = \frac{D-2}{24} - \frac{26- D}{48} \left( \left( \frac{J_{1,2}}{J_{3,4}} \right)^{\frac{1}{4}} - \left( \frac{J_{3,4}}{J_{1,2}} \right)^{\frac{1}{4}} \right)^2
\end{align}
for open and closed Nambu-Goto strings. Here $J_{1,2}$ and $J_{3,4}$ are angular momenta in the $1-2$ and $3-4$ planes in which the closed string rotates.
Regarding this result, two aspects are puzzling:
\begin{itemize}
\item For the covariantly quantized string, $a$ is a constant which is independent of the ratio of the angular momentum components, in contrast to \eqref{eq:a_HS} for $D \neq 26$.
\item In the limit $J_{3,4} \to 0$, the closed string degenerates to two straight open strings with the same endpoints. One would thus expect $a_{\mathrm{open}} = a_{\mathrm{closed}}$, which, however, 
only holds for $D = 26$. Even worse, for $D < 26$, $a_{\mathrm{closed}}$ diverges in the limit $J_{3,4} \to 0$.
\end{itemize}

Due to this seeming inconsistency in non-critical dimension, it seems desirable to verify \eqref{eq:a_HS} with other methods. The result for the open string, including a generalization to masses at the endpoints, was recently re-derived in \cite{RotatingStringEnergy} using a semi-classical approximation, i.e., by quantizing the perturbations around classical rotating string solutions, and using a local renormalization procedure to extract quantum corrections to the energy. Here, we perform the analogous analysis for closed strings, also verifying \eqref{eq:a_HS}.

The semi-classical quantization procedure in principle also allows to compute the spectrum of excitations of the classical rotating solutions. In \cite{RotatingStringEnergy} it was shown that for the open string (at least for the low-lying excitations) the semi-classical approach leads to the same spectrum of physical excitations as the covariant quantization scheme. Here, we show that for the closed string the spectrum of excitations in the semi-classical theory is too large, i.e., there are semi-classical excitations that do not correspond to physical excitations of the full non-perturbative covariantly quantized theory. That such phenomena are possible in gauge theories is well known: For example, already at the classical level, Yang-Mills theory is not linearization stable \cite{Arms81}, i.e., solutions to the linearized equations are not necessarily tangent to the manifold of solutions of the full equations, i.e., do not derive from a curve in the set of solutions of the full equations.   More generally, denoting by $s_0$ the linearized part of the BRST operator $s$, the cohomology $H(s_0)$ describing observables in the linearized theory is in general larger than the cohomology $H(s)$ of the full theory, an elementary example being again provided by Yang-Mills theory, \cf \cite{BrandtDragonKreuzer89} for example. In the present case, the reason seems to be that the level-matching condition is not implemented at the linearized level.

Our perturbative approach is based on the finding \cite{BRZ} that in the sense of perturbation theory around non-degenerate classical solutions, the Nambu-Goto string can be quantized for any $D$, without anomalies. Let us sketch the argument: Let $X: \Sigma \to (\R^D, \eta)$ with $\eta = \diag(- + \dots +)$ be the embedding of the world-sheet $\Sigma$ into the target Minkowski space and split $X$ into a classical solution $\bar X$ and a perturbation $\vp$, i.e.,
\beq
\label{eq:X_phi}
 X = \bar X + \gamma^{-\frac{1}{2}} \vp,
\eeq 
where $\gamma^{-\frac{1}{2}}$ is seen as a formal expansion parameter.
The Nambu-Goto action is 
\beq
\label{eq:S_NG}
 \mathcal{S} = - \gamma \int_\Sigma \sqrt{g} \ud^2 x
\eeq
where $g$ is the determinant of the metric induced by the embedding. The full action is invariant under reparametrizations of embedding, giving rise to the BRST transformation of $\vp$ given by
\beq
\label{eq:s_vp}
 s \vp^a = c^\mu \del_\mu X^a = c^\mu \del_\mu \bar X^a + \gamma^{-\frac{1}{2}} c^\mu \del_\mu \vp^a,
\eeq
with $c^\mu$ the diffeomorphism (re-parametrization) ghost vector field. Here roman (greek) indices are target space (world-sheet) indices. 
Given a gauge condition $T_\mu[\vp] = 0$, one supplements the action \eqref{eq:S_NG} by the gauge-fixing action
\beq
 \mathcal{S}_{\mathrm{gf}} = \int_\Sigma \left( b^\mu T_\mu[\vp] - \bar c^\mu s T_\mu[\vp] \right) \sqrt{\bar g} \ud^2 x,
\eeq
with $\bar g$ the determinant of the metric $\bar g_{\mu \nu}$ induced by $\bar X$. This action is invariant under the BRST transformation defined by \eqref{eq:s_vp} and
\beq
 s c^\mu = \gamma^{-\frac{1}{2}} c^\nu \bar \nabla_\nu c^\mu, \qquad s \bar c^\mu = b^\mu, \qquad s b^\mu = 0,
\eeq
where $\bar \nabla$ is the Levi-Civita covariant derivative \wrt $\bar g_{\mu \nu}$.

The possible occurrence of anomalies, such as a violation of the nilpotency of the BRST charge, is governed by the cohomologies of $s$ and $s$ mod $\ud$. The free, i.e., $\order(\gamma^0)$, part of $s$ is
\beq
\label{eq:s_0}
 s_0 \vp^a = c^\mu \del_\mu \bar X^a.
\eeq
For a non-degenerate classical embedding $\bar X$, $\del_\mu \bar X^a$ has maximal rank. It follows that the fluctuations $\vp^a$ tangential to the embedding $\bar X$ form trivial pairs with the ghosts $c^\mu$ and drop out of the cohomology of $s_0$. Hence, $s_0$ has trivial cohomology at positive ghost number. By cohomological perturbation theory \cite{PiguetSorella}, this carries over to the cohomology of $s$, and also to the cohomology of $s$ mod $\ud$. Hence, there is a renormalization scheme in which no anomalies are present, in any order of perturbation theory. 

In the following, we will truncate the action at $\order(\gamma^0)$, i.e., at second order in the dynamical fields $\vp^a$, $c_\mu$, $\bar c_\mu$, $b_\mu$. This free action is the one relevant for the semi-classical approximation. Furthermore, we use transversal gauge \cite{Kleinert86}, where the absence of anomalies at the level of the free theory is manifest. For this, one chooses
\beq
\label{eq:GF}
 T_\mu[\vp] = \del_\mu \bar X^a \vp_a.
\eeq
By the equations of motion derived from the free action, the unphysical fields, i.e., the auxiliary fields $c_\mu$, $\bar c_\mu$, $b_\mu$ and the pure gauge perturbations, i.e., those parallel to $\del_\mu \bar X^a$, then vanish,\footnote{At second order in the perturbation, the fluctuations parallel to the classical embedding $\bar X$ drop out of the (not yet gauge fixed) Nambu-Goto action \cite{BRZ}. From this, one easily derives that in transversal gauge the unphysical fields vanish on-shell at the level of the free theory.} so they can be consistently set to zero at the level of the free theory.\footnote{The possibility of such a trivial gauge fixing (and also the triviality of the cohomology of $s$)  is due to the absence of derivatives of the ghost in the BRST transformation \eqref{eq:s_vp}. This feature in particular distinguishes the model from Yang-Mills theories and General Relativity.} In particular, the corresponding free BRST charge vanishes\footnote{As a consequence of the absence of derivatives of dynamical fields in the gauge fixing condition \eqref{eq:GF}, the free BRST current, i.e., the Noether current derived from $s_0$ and the free part of the action, vanishes. (The Nambu-Goto action $\mathcal{S}$ does not contribute to the free BRST current \cite{BRZ}.)} (and is thus also nilpotent). Hence, expanding the action to $\mathcal{O}(\gamma^0)$ and restricting to fluctuations $\vp^a$ normal to the classical embedding $\bar X$ (these vanish under the action of $s_0$), one arrives at a free theory involving only physical fields.


The perturbation $\vp$ is naturally interpreted as a field living on the world-sheet, so that quantization yields a free quantum theory on the world-sheet.
The equations of motion for $\vp$
only depend on the world-sheet geometric data, i.e., the metric and the second fundamental form induced by the classical embedding $\bar X$. Hence, it seems natural, in line with the framework of \cite{BRZ}, to use methods from quantum field theory on curved space-time \cite{HollandsWaldWick, HollandsWaldReview} for the renormalization of the free world-sheet Hamiltonian $H^0$. This means that renormalization is performed locally, using the local geometric data. The correspondence between the world-sheet Hamiltonian and the target space energy then gives corrections to the classical Regge trajectories.

Let us analyze the latter point in more detail. We recall that in the covariant quantization the ground state energies $E_{J}$ for angular momentum $J_{1,2}, J_{3,4} \geq 0$ 
satisfy
\beq
\label{eq:ReggeTrajectory}
 E^2 = 4 \pi \gamma ( J_{1,2} + J_{3,4} - 2 a),
\eeq
with $\gamma$ the string tension, a relation which we call the Regge trajectory in the following (no extension of the angular momenta to the complex plane is intended). It is important to note that this ground state is degenerate in the sense that ground states for fixed $J = J_{1,2} + J_{3,4}$ all have the same energy $E_J$. We will be parameterizing the rotating string in terms of the two radii $R_{1,2}$, $R_{3,4}$ in the two planes. In terms of these, the 
classical target space energy and angular momentum are given by
\begin{align}
\label{eq:E_class}
 \bar E & = 2 \pi \gamma R, &
 \bar J_{1, 2} & = \pi \gamma R_{1,2}^2, &
 \bar J_{3, 4} & = \pi \gamma R_{3,4}^2,
\end{align}
with
\beq
\label{eq:Def_R}
 R = \sqrt{ R_{1,2}^2 + R_{3,4}^2 }.
\eeq
Here the bar stands for the values corresponding to the classical (background) configuration $\bar X$, \cf \eqref{eq:X_phi}.
In our parametrization, the relation between the world-sheet Hamiltonian $H$, the quantum correction $E^q$ to the target space energy $E$, and the quantum corrections $J^q_{1,2}$, $J^q_{3,4}$ to the angular momenta $J_{1,2}$, $J_{3,4}$ is
\beq
\label{eq:Relation_H_E}
 E^q = \frac{1}{R} \left( H + J^q_{1,2} + J^q_{3,4} \right),
\eeq 
yielding
\begin{align}
 E^2 & = \left( \bar E + E^q \right)^2 \nn \\
 & = 4 \pi^2 \gamma^2 R^2 + 4 \pi \gamma \left( H + J^q_{1,2} + J^q_{3,4} \right) + \order(R^{-2}) \nn \\
\label{eq:E2}
 & = 4 \pi \gamma \left( J_{1, 2} + J_{3,4} + H^0 \right) + \order(R^{-1}).
\end{align}
By comparison with \eqref{eq:ReggeTrajectory}, one can directly read off the intercept $a$ from the expectation value of $H^0$, i.e.,
\beq
\label{eq:InterceptDetermination}
 a = - \tfrac{1}{2} \VEV{H^0}.
\eeq

As the case of elliptic strings is notationally and computationally much more involved, we chose to first present our calculation for the case of circular strings, i.e., for $R_{1,2}=R_{3,4}$,
and to discuss the modifications necessary for the treatment of elliptic strings in a separate section. The article is thus structured as follows: In the next section, we discuss the perturbations of classical rotating circular solutions, and in Section~\ref{sec:Energy} the calculation of the corresponding intercept. Section~\ref{sec:Elliptic} deals with the generalization to the elliptic case. These calculations are for simplicity performed for $D=6$. The straightforward generalization to general $D$ is discussed in Section~\ref{sec:D}. Finally, in Section~\ref{sec:ExcitationSpectrum}, we compare the spectrum of physical excitations in the linearized semi-classical theory and the covariant quantization scheme.
We conclude in Section~\ref{sec:Conclusion}. An appendix contains some intermediate results of our treatment of the elliptic case.

\section{Perturbations of classical rotating circular strings}
\label{sec:ClassicalSolutions}


For an embedding $X: \Sigma \to (\R^D, \eta)$ the Nambu-Goto action \eqref{eq:S_NG} yields the equations of motion
\beq
\label{eq:eomX}
 \Box_g X  = 0.
\eeq
We also recall the target space momentum and angular momentum derived from the action \eqref{eq:S_NG}:
\begin{subequations}
\label{subeq:P_L}
\begin{align}
\label{eq:Momentum}
 P^i & = \int_0^{2 \pi} \frac{\delta \mathcal{S}}{\delta \del_0 X_i} \ud \sigma = - \gamma \int_0^{2 \pi} \sqrt{g} g^{0 \nu} \del_\nu X^i \ud \sigma, \\
\label{eq:AngularMomentum}
 L_{ij} & = \int_0^{2 \pi} \left[ \frac{\delta \mathcal{S}}{\delta \del_0 X^j} X_i - i \leftrightarrow j \right] \ud \sigma = \gamma \int_0^{2 \pi} \sqrt{g} g^{0 \nu} \left( X_j \del_\nu X_i - X_i \del_\nu X_j \right) \ud \sigma.
\end{align}
\end{subequations}
Here we assumed $\Sigma$ to be parameterized by $(\tau, \sigma) \in \R \times [0, 2 \pi)$. The target space energy is given by $E = P^0$.

We parameterize the rotating circular solution as
\begin{equation}
\label{eq:X}
 \bar X(\tau, \sigma) = \frac{R}{\sqrt 2} \begin{pmatrix} \sqrt{2} \tau \\ \sin \tau \cos \sigma \\ - \cos \tau \cos \sigma \\ \cos \tau \sin \sigma \\ \sin \tau \sin \sigma \\ 0 \end{pmatrix},
\end{equation}
where $\sigma \in [0, 2 \pi)$ and $R$ is given by \eqref{eq:Def_R}. Note that in our parametrization, the world sheet coordinates $\tau, \sigma$ are dimensionless.
For simplicity, we here assumed that the target space-time is six dimensional. 
As discussed in Section~\ref{sec:D}, it is straightforward to add further dimensions (or to remove the sixth).
The induced metric on the world-sheet, in the coordinates introduced above, is a multiple of the Minkowski metric,
\beq
\label{eq:metric}
 \bar g_{\mu \nu} = \frac{R^2}{2} \eta_{\mu \nu}.
\eeq
In particular, the equation of motion \eqref{eq:eomX} can be easily checked. Energy and angular momentum of the above solution were given in \eqref{eq:E_class}, with $R_{1,2} = R_{3,4} = \frac{R}{\sqrt{2}}$.

Our goal is to perform a (canonical) quantization of the fluctuations $\vp$ around the classical background $\bar X$, \cf \eqref{eq:X_phi}. As discussed in the introduction, we may, at the level of the free theory, restrict to normal fluctuations. We parameterize these as
\begin{align}
\label{eq:phi_Parametrization}
 \varphi & = f_s v_s + f_a v_a + f_b v_b + f_c v_c,
\end{align}
with $v_s$, $v_a$, $v_b$, $v_c$, orthonormal and normal to the world-sheet. Concretely, we choose
\begin{align}
\label{eq:Def_v}
 v_s & = \begin{pmatrix} 0 \\ 0 \\ 0 \\ 0 \\ 0 \\ 1 \end{pmatrix}, &
 v_a & = \sqrt{2} \begin{pmatrix} \frac{1}{\sqrt{2}} \\ \cos \tau \cos \sigma \\ \sin \tau \cos \sigma \\ - \sin \tau \sin \sigma \\ \cos \tau \sin \sigma \\ 0 \end{pmatrix}, &
 v_b & = \begin{pmatrix} 0 \\ \sin \tau \cos \sigma \\ - \cos \tau \cos \sigma \\ \cos \tau \sin \sigma \\ \sin \tau \sin \sigma \\ 0 \end{pmatrix}, &
 v_c & =\begin{pmatrix} 0 \\ \cos \tau \sin \sigma \\ \sin \tau \sin \sigma \\ \sin \tau \cos \sigma \\ - \cos \tau \cos \sigma \\ 0 \end{pmatrix}.
\end{align}
Here the \emph{scalar} component $f_s$ describes the fluctuations in the direction perpendicular to the planes of rotation. The mode $f_a$ describes infinitesimal rotations generated by $L_{1,2} + L_{3,4}$ ($v_a$ also has a time component, which is present to make it orthogonal to the world-sheet). The mode $f_b$ describes dilations in the subspace spanned by $e_1$, $e_2$, $e_3$, $e_4$ and $f_c$ describes combined shears in the planes spanned by $e_1$, $e_3$ and $e_2$, $e_4$. 
Subsuming the modes in the planes of rotation into $f = (f_a, f_b, f_c)$, we obtain, at $\order(\gamma^0)$, the free action
\begin{equation}
\label{eq:S0}
 \sS^0 = \frac{1}{2} \int \left( {\dot f}^2 - {f'}^2 - f^T A \dot f - f^T B f' - f^T C f + {\dot f_s}^2 - {f'_s}^2 \right) \ud \sigma \ud \tau,
\end{equation}
where $f^T$ is the transpose of $f$, derivates \wrt $\tau$ are denoted by dots and those \wrt $\sigma$ by primes and
\begin{align}
 A & = \begin{pmatrix} 0 & \sqrt{8} & 0 \\ -\sqrt{8} & 0 & 0 \\ 0 & 0 & 0 \end{pmatrix}, &
 B & = \begin{pmatrix} 0 & 0 & - \sqrt{8} \\ 0 & 0 & 0 \\ \sqrt{8} & 0 & 0 \end{pmatrix}, &
 C & = \begin{pmatrix} 0 & 0 & 0 \\ 0 & -4 & 0 \\ 0 & 0 & 4 \end{pmatrix}.
\end{align}
We see that the scalar field decouples. 


From \eqref{eq:S0}, we see that the scalar fluctuations decouple from the others, grouped into $f$, termed the \emph{planar sector} in the following. 
We obtain the equations of motion
\begin{subequations}
\begin{align}
\label{eq:eom_s}
 0 & = - \ddot f_s + f''_s, \\
\label{eq:eom_p}
 0 & = - \ddot f + f'' - A \dot f - B f' - C f.
\end{align}
\end{subequations}
It is easily checked that the symplectic forms for the scalar and the planar sector are
\beq
\label{eq:SymplecticForm}
\sigma(f_s, \tilde f_s) = \int_{\Sigma_\tau} ( f_s \dot{\tilde f}_s - \dot f_s \tilde f_s)  \ud \sigma, \qquad
 \sigma(f, \tilde f) = \int_{\Sigma_\tau} \begin{pmatrix} f & \dot f \end{pmatrix}
   \begin{pmatrix}
 A & \1_3 \\
 - \1_3 & 0_3 
\end{pmatrix}
  \begin{pmatrix} \tilde f \\ \dot{\tilde f} \end{pmatrix} \ud \sigma
\eeq
where $\Sigma_\tau$ is the time-slice at an arbitrary time $\tau$.

In order to prepare for a canonical quantization, we identify symplectically normalized mode solutions.\footnote{For an exposition of canonical quantization in the presence of single time derivatives in the action, we refer to the appendix of \cite{Fulling}. Our results, in particular regarding the behavior of modes that degenerate in the symplectic form, are in complete agreement with the general properties derived there.}
For the scalar component, one finds the solutions
\beq
\label{eq:f_s_n}
 f_{s, n}^\pm = \tfrac{1}{\sqrt{4 \pi n}} e^{\pm i n \sigma} e^{- i n \tau} 
\eeq
for $n \geq 1$. These are normalized \wrt their symplectic form, i.e.,
\beq
\label{eq:SymplecticNormalization}
 \sigma(\overline{f_{s, n}^\alpha}, f_{s, m}^{\beta}) = - i \delta_{m n} \delta^{\alpha \beta},
\eeq
where the bar stands for complex conjugation and $\alpha, \beta = \pm$.
There is also a zero mode $f_q$ which degenerates in the symplectic form and can not be normalized. It forms a canonical pair with a supplementary linearly growing mode $f_p$, i.e.,
\begin{align}
\label{eq:f_q_f_p}
 f_q & = \tfrac{1}{\sqrt{2 \pi}}, &
 f_p & = \tfrac{1}{\sqrt{2 \pi}} \tau,
\end{align}
with $\sigma(f_q, f_p) = 1$. These correspond to position and momentum in the direction perpendicular to the planes of rotation, and can be seen as the Nambu-Goldstone modes for the broken invariance under translations perpendicular to the planes of rotation. In the determination of the Regge intercept, these will be discarded, as we have to fix the momentum.

For the planar sector, we find the positive energy solutions
\begin{subequations}
\begin{align}
\label{eq:f_1_n}
 f_{1, n}^\pm & = \tfrac{1}{\sqrt{8 \pi n (n^2 - 1)}} (\pm i \sqrt{2}, \mp n, n) e^{\pm i n \sigma} e^{- i n \tau}, \\
\label{eq:f_2_n}
 f_{2, n}^\pm & = \tfrac{1}{\sqrt{16 \pi (n - 1)}} (\mp i \sqrt{2}, \pm 1, 1) e^{\pm i (n-2) \sigma} e^{- i n \tau}, \\
\label{eq:f_3_n}
 f_{3, n}^\pm & = \tfrac{1}{\sqrt{16 \pi (n + 1)}} (\pm i \sqrt{2}, \pm 1, 1) e^{\pm i (n+2) \sigma} e^{- i n \tau}.
\end{align}
\end{subequations}
These are again normalized \wrt the symplectic form, analogously to \eqref{eq:SymplecticNormalization}, with a few exceptions, which we now discuss in detail.

For $n=0$, the modes $f_1^\pm$ degenerate to a single mode, which is degenerate in the symplectic form. There is another linearly growing solution complementing it, so that
\begin{subequations}
\begin{align}
\label{eq:f_theta}
 f_\theta & = \tfrac{1}{\sqrt{2 \pi}} (1, 0, 0), \\
\label{eq:f_lambda}
 f_\lambda & = \tfrac{1}{\sqrt{2 \pi}} \left[ - \tau (1, 0, 0) + \tfrac{1}{\sqrt{2}} (0, 1, 0) \right],
\end{align}
\end{subequations}
form a canonical pair,
\beq
\label{eq:sigma_theta_lambda}
 \sigma(f_\theta, f_\lambda) = 1.
\eeq
Recalling that $f_a$ could be identified as an infinitesimal rotation generated by $L_{1,2} + L_{3, 4}$, we can see these as Goldstone modes for the broken $L_{1,2} + L_{3, 4}$ invariance. We should think of $f_\theta$ as an angle and $f_\lambda$ as the corresponding angular momentum variable.\footnote{However, we draw attention to the sign of the term linear in $\tau$ in \eqref{eq:f_lambda} compared to \eqref{eq:f_theta}, which was introduced in order to have usual sign in \eqref{eq:sigma_theta_lambda}. The same phenomenon occurs for the open string, and in fact also for an analogous semi-classical treatment of the hydrogen atom \cite{RotatingStringEnergy}.} This is further supported below. Furthermore, as shown in the next section, these modes are not relevant for the determination of the Regge intercept.

The modes $f_2$ and $f_3$ do not degenerate in the symplectic form for $n=0$, so they are not conventional null modes. However, $f_{2,0}^\pm$ does not fulfill the normalization \eqref{eq:SymplecticNormalization} (the sign is changed). It should thus be interpreted as a negative energy mode. Note that $\overline{f_{2,0}^\pm} = i f^\pm_{3,0}$, so $f^\pm_{3, 0}$ and $i f^\pm_{2, 0}$ are symplectic pairs of a positive and a negative energy solution.
As $L_{1,3} + L_{2, 4}$ and $L_{1,2} - L_{3, 4}$ both commute with $L_{1,2} + L_{3, 4}$, it is suggestive to regard the modes $f^\pm_{3,0}$ as the generators of these rotations.\footnote{Also $L_{1,4} - L_{2, 3}$ commutes with $L_{1,2} + L_{3, 4}$, but it generates re-parametrizations.} This is explicitly checked below.
Also note that ``positive energy'' might be a misnomer, as one can easily check that these modes do not contribute to the free Hamiltonian, see below.
It follows that these modes lead to a degeneracy of the ground state. 
This reflects the classical ground state degeneracy discussed in the introduction.

For $n=1$, the modes $f_1$ and $f_2$ degenerate in the symplectic form and can thus 
not be normalized. Omitting the normalization, $f^\pm_{1,1}$ and $f^\pm_{2, 1}$ are solutions and 
 $f^\mp_{1, 1}$ and $f^\pm_{2, 1}$ coincide. It turns out that there are linearly growing solutions complementing these. We thus have
\begin{subequations}
\begin{align}
 f_q^\pm & = \tfrac{1}{\sqrt{8 \pi}} (\pm i \sqrt{2}, \mp 1, 1) e^{\pm i \sigma} e^{- i \tau}, \\
 f_p^\pm & = \tfrac{1}{\sqrt{8 \pi}} \left[ \tau (\pm i \sqrt{2}, \mp 1, 1) + \tfrac{i}{8} (\mp i 3 \sqrt{2}, \mp 1, 1) \right] e^{\pm i \sigma} e^{- i \tau}.
\end{align}
\end{subequations}
Together with their complex conjugates, these form canonical pairs,
\begin{align}
 \sigma(\overline{f_q^\pm}, f_p^\pm) & = 1, &
 \sigma(f_q^\pm, \overline{f_p^\pm}) & = 1, 
\end{align}
with all other combinations vanishing. Hence, $f_q^\pm, \overline{f_q^\pm}$ can be interpreted as positions and $\overline{f_p^\pm}, f_p^\pm$ as the corresponding momenta. We thus have four canonical pairs of position and momenta. These have the natural interpretation as center-of-mass positions and momenta in the sub-space spanned by $e_1$, $e_2$, $e_3$, $e_4$, i.e., as the Nambu-Goldstone modes for the broken invariance under translations in this subspace.
Similarly to the scalar null modes $f_q, f_p$, these will be discarded in the determination of the Regge intercept.

An analogous identification of these special modes for the open string was made in \cite{RotatingString, RotatingStringEnergy}. It can also be checked explicitly: Writing
\begin{subequations}
\label{subeq:phi_Expansion}
\begin{align}
\label{eq:phi_Expansion}
 f & = \sum_{\alpha \in \pm} \left[ \sum_{r \in \{ 1, 2, 3 \}} \sum_{n \in \N_r} \left( a_{r, n}^\alpha f^\alpha_{r, n} \right) + \left( q^\alpha f_q^\alpha + p^\alpha f_p^\alpha \right) + \text{h.c.} \right] + \theta f_\theta + \lambda f_\lambda, \\
\label{eq:phi_s_Expansion}
 f_s & = \sum_{n \in \N_s} \sum_{\alpha \in \pm} \left( a_{s, n}^\alpha f^\alpha_{s, n} + \text{h.c.} \right) + q f_q + p f_p,
\end{align}
\end{subequations}
where
\begin{align}
 \N_s & = \{ n \geq 1 \}, &
 \N_1 & = \N_2 = \{ n \geq 2 \}, &
 \N_3 & = \{ n \geq 0 \},
\end{align}
and the coefficients $q, p, \theta, \lambda$ are real, one finds, for the expansion of the energy, angular momenta, and momenta, \cf \eqref{subeq:P_L},
\begin{subequations}
\begin{align}
\label{eq:E_1st_order}
 E & = \bar E + \sqrt{2 \pi \gamma} \lambda + \order(\gamma^0), \\
\label{eq:L_1st_order}
 L_{1, 2} + L_{3, 4} & = \bar L_{1, 2} + \bar L_{3, 4} + \sqrt{2 \pi \gamma} R \lambda + \order(\gamma^0), \\
 L_{1, 2} - L_{3, 4} & =  \sqrt{2 \pi \gamma} R \Re (a^+_{3,0} - a^-_{3,0} ) + \order(\gamma^0), \\
 L_{1, 3} + L_{2, 4} & =  \sqrt{2 \pi \gamma} R \Im (a^+_{3,0} + a^-_{3,0} ) + \order(\gamma^0), \\
\label{eq:P1_1st_order}
 P^{1/3} & = \sqrt{2 \pi \gamma} \Im ( \mp p^+ + p^- ) + \order(\gamma^0), \\
\label{eq:P2_1st_order}
 P^{2/4} & = \sqrt{2 \pi \gamma} \Re ( \pm p^+ - p^- ) + \order(\gamma^0), \\
\label{eq:P5_1st_order}
 P^5 & = \sqrt{2 \pi \gamma} p + \order(\gamma^0).
\end{align}
\end{subequations}
This supports the identification of the modes $f_p$, $f_\lambda$, $f_p^\pm$, $f^\pm_{3,0}$ with (angular) momenta discussed above.

Canonical quantization now proceeds as follows: One introduces annihilation and creation operators $\hat a^\pm_{r, n}$, $\hat{a}^{\pm *}_{r, n}$ for $r \in \{ s, 1, 2, 3 \}$, $n \in \N_r$, fulfilling
\beq
 [ \hat a^\alpha_{r, n}, \hat a^{\beta *}_{r', n'} ] = \delta^{\alpha \beta} \delta_{r r'} \delta_{n n'}.
\eeq
Furthermore, one introduces position operators $\hat q, \hat \theta, \hat q^\pm, \hat q^{\pm *}$ and momenta $\hat p, \hat \lambda, \hat p^\pm, \hat p^{\pm *}$ with commutation relations
\begin{align}
 [\hat q, \hat p] & = i, &
 [\hat \theta, \hat \lambda] & = i, &
 [\hat q^{\alpha *}, \hat p^\beta] & = i \delta^{\alpha \beta}, &
 [\hat q^\alpha, \hat p^\beta] & = 0.
\end{align}
The complex positions $\hat q^\pm$ and momenta $\hat p^\pm$ can be represented on $L^2(\R^2)$ with canonical position and momentum operators $\hat q^\pm_i$, $\hat p^\pm_i$ as
\begin{align}
\label{eq:qp}
 \hat q^\pm & = \tfrac{1}{\sqrt{2}} ( \hat q_1^\pm + i \hat q_2^\pm ), &
 \hat p^\pm & = \tfrac{1}{\sqrt{2}} ( \hat p_1^\pm + i \hat p_2^\pm ).
\end{align}
In particular, with \eqref{eq:P1_1st_order}, \eqref{eq:P2_1st_order}, this implies
\begin{subequations}
\label{subeq:P}
\begin{align}
\label{eq:P_p_i_1}
 P^{1/3} & = \sqrt{\pi \gamma} \left( \mp \hat p_2^+ + \hat p_2^- \right) + \order(\gamma^0), \\
\label{eq:P_p_i_2}
 P^{2/4} & = \sqrt{\pi \gamma} \left( \pm \hat p_1^+ - \hat p_1^- \right) + \order(\gamma^0).
\end{align}
\end{subequations}
The canonically conjugate positions are thus
\begin{subequations}
\label{subeq:X}
\begin{align}
\label{eq:X_q_i_1}
 X^{1/3} & = \left( 2\sqrt{\pi \gamma} \right)^{-1} \left( \mp \hat q_2^+ + \hat q_2^- \right) + \order(\gamma^{-1}), \\
\label{eq:X_q_i_2}
 X^{2/4} & = \left( 2\sqrt{\pi \gamma} \right)^{-1} \left( \pm \hat q_1^+ - \hat q_1^- \right) + \order(\gamma^{-1}).
\end{align}
\end{subequations}
That these are the correct center of mass (cms) positions can indeed be explicitly checked. For example, up to a reparametrization, the combination $\sqrt{2 \pi \gamma} \Im (f^+_q - f^-_q)$ parameterizes a shift by one unit in direction $e_1$.

Canonical quantization then proceeds by replacing the coefficients in \eqref{subeq:phi_Expansion} by the hatted corresponding operators. 
One can then explicitly check the fulfilment of the canonical equal time commutation relations, as dictated by the symplectic form \eqref{eq:SymplecticForm}.
Omitting the positions and momenta (this will be justified below) we thus have quantum fields $\hat f_s$, $\hat f$ with two-point functions
\begin{subequations}
\label{subeq:2pt}
\begin{align}
\label{eq:2pt_s}
 w_s(x; x') & \defeq \bra{\Omega} \hat f_s(x) \hat f_s(x') \ket{\Omega} = \sum_{n \in \N_s} \sum_{\alpha \in \pm} f^\alpha_{s, n}(x) \overline{f^\alpha_{s, n}}(x'), \\
\label{eq:2pt_p}
 w_{p, ij}(x; x') & \defeq \bra{\Omega} \hat f_i(x) \hat f_i(x') \ket{\Omega} = \sum_{k = 1}^3 \sum_{n \in \N_k} \sum_{\alpha \in \pm} f^\alpha_{k, n, i}(x) \overline{f^\alpha_{k, n, j}}(x'),
\end{align}
\end{subequations}
where in the second equation $i, j$ run over the labels $(a, b, c)$.

\section{The world-sheet Hamiltonian}
\label{sec:Energy}

The free Hamiltonian corresponding to the free action~\eqref{eq:S0} is
\begin{equation}
\label{eq:H_0}
 H^0 = \frac{1}{2} \int_{0}^{2 \pi} \left( \dot f^2 + {f'}^2 + f^T C f + f^T B f' + \dot f_s^2 + {f'_s}^2 \right) \ud \sigma.
\end{equation}
In terms of the creation and annihilation and position and momentum operators introduced in the previous section, this Hamiltonian formally reads as
\begin{equation}
\label{eq:H_0_ladder}
 H^0 = \frac{1}{2} \left( \sum_{r \in \{ s, 1, 2, 3 \}} \sum_{n \in \N_r} \sum_{\alpha \in \pm} n \left( \hat a^\alpha_{r, n} \hat a^{\alpha *}_{r, n} + \hat a^{\alpha *}_{r, n} \hat a^\alpha_{r, n} \right) + \hat p^2 - \hat \lambda^2 \right) + \sum_{\alpha \in \pm} \left( \hat p^{\alpha *} \hat p^\alpha + i \hat p^\alpha \hat q^{\alpha *} - i \hat p^{\alpha *} \hat q^\alpha \right).
\end{equation}

Rewriting the last term in \eqref{eq:H_0_ladder} in terms of the position and momentum operators introduced in \eqref{eq:qp}, one obtains
\beq
 \sum_{\alpha \in \pm} \left( \hat p^{\alpha *} \hat p^\alpha + i \hat p^\alpha \hat q^{\alpha *} - i \hat p^{\alpha *} \hat q^\alpha \right) = \sum_{\alpha \in \pm} \left[ \tfrac{1}{2} \left( (\hat p^\alpha_1)^2 + (\hat p^\alpha_2)^2 \right) + \hat p^\alpha_1 \hat q^\alpha_2 - \hat p^\alpha_2 \hat q^\alpha_1 \right].
\eeq
By \eqref{subeq:P}, \eqref{subeq:X}, the last two terms are precisely the leading order term of the cms contribution to the angular momentum $-L_{1,2} - L_{3,4}$. As a cms contribution to the angular momentum does not change the energy (for a given momentum), this term must be present in $H^0$, due to \eqref{eq:Relation_H_E}. For the determination of the Regge trajectory, such a cms contribution should of course be absent. Hence, we choose a state in which the expectation value of the last two terms in \eqref{eq:H_0_ladder} vanishes.

To evaluate the rest mass, we should work at vanishing spatial momentum. Obviously, we may choose a state in which the expectation values of $(\hat p^\alpha_1)^2 + (\hat p^\alpha_2)^2$ and $\hat p^2$ are arbitrarily small. We can hence neglect the momenta in \eqref{eq:H_0_ladder}.\footnote{As a consistency check, we note that the leading order contribution to $E^2$ of the terms quadratic in the momenta is given by $2 \pi \gamma (\hat p^2 + \sum_{\alpha, i} (\hat p_i^\alpha)^2 )$, \cf \eqref{eq:E2}. With \eqref{eq:P1_1st_order}, \eqref{eq:P2_1st_order}, and \eqref{eq:P5_1st_order}, we thus see that the momenta contribute $(P^i)^2$ to $E^2$, up to corrections of $\order(\gamma^{-\frac{1}{2}})$, as expected.} 

We now argue that also the $\hat \lambda^2$ term in \eqref{eq:H_0_ladder} should be neglected. First, we recall that in \eqref{eq:Relation_H_E}, the quantum corrections $E^q$, $L^q_{1,2}$, $L^q_{3,4}$ are obtained by expanding the classical expressions \eqref{eq:Momentum} for the target space energy $E = P^0$ and \eqref{eq:AngularMomentum} for the angular momentum in $\vp$, neglecting the zeroth order (classical) terms. The world-sheet Hamiltonian $H$ generates (world-sheet) time translations on the world-sheet. The quantum target space energy $E^q$, should generate time translations in target space. However, the background world-sheet is rotating in the $1$-$2$ and the $3$-$4$ plane and so are the vectors $v_a$, $v_b$, $v_c$ parameterizing the fluctuations. The correct relation between $H$ and $E^q$ is thus \eqref{eq:Relation_H_E}, where the factor $\frac{1}{R}$ corrects the different scaling of target space time $X^0$ and world-sheet time $\tau$, and $L^q_{1,2} + L^q_{3,4}$ corrects for the missing rotation. One can easily check \eqref{eq:Relation_H_E} explicitly up to $\order(\vp^2)$, which is the relevant order for our purposes. A further relation between the quantum target space energy $E^q$ and the world-sheet Hamiltonian $H$ is
\beq
 E^q = \frac{1}{R} H \mod \frac{1}{R},
\eeq
which follows from the fact that a world-sheet translation $\tau \mapsto \tau + 2 \pi$ corresponds to a target space translation $X^0 \mapsto X^0 + 2 \pi R$. Comparison with \eqref{eq:Relation_H_E} shows that the spectrum of $L^q_{1,2} + L^q_{3,4}$ should be discrete,
\beq
 L^q_{1,2} + L^q_{3,4} \in \Z.
\eeq
This of course matches the expectation for the spectrum of (the sum of) angular momentum operators. Note that this is a non-perturbative statement, which can not be deduced from the expansion of $L^q_{1,2} + L^q_{3,4}$ in $\vp$. Due to \eqref{eq:L_1st_order}, this means that the spectrum of $\hat \lambda$ is discrete. In particular, we may choose the eigenstate of eigenvalue $0$, corresponding to fixing the angular momentum $L_{1,2} + L_{3,4}$ to the classical value, up to corrections of $\order(\gamma^0)$. By \eqref{eq:E_1st_order}, this is also an eigenstate of $E$, again up to corrections of $\order(\gamma^0)$.

We now come to the evaluation of the expectation value of $H^0$ (with the (angular) momenta set to zero). As in \cite{RotatingStringEnergy}, we follow the proposal \cite{BRZ} to 
employ a locally covariant renormalization technique \cite{HollandsWaldWick} developed for QFT on curved space-times.
Concretely, the expectation value of Wick squares (possibly with derivatives) is determined as follows:
\beq
 \bra{\Omega} (\nabla^\alpha \phi \nabla^\beta \phi)(x) \ket{\Omega} = \lim_{x' \to x} \nabla^\alpha {\nabla'}^\beta \left( w(x; x') - h(x; x') \right)
\eeq
Here $\alpha, \beta$ are multiindices, $w$ is the two-point function in the state $\Omega$, defined as on the \lhs of \eqref{subeq:2pt}, and $h$ is a distribution which is covariantly constructed out of the geometric data, the \emph{Hadamard parametrix}. For physically reasonable states, the difference $w - h$ is smooth, so that the above coinciding point limit exists and is independent of the direction from which $x'$ approaches $x$.

For our purposes, it is advantageous to perform the limit of coinciding points from the time direction, i.e., we take $x = (\tau, \sigma)$, $x' = (\tau + t, \sigma)$, and $t \to +0$. Performing the summation in \eqref{eq:2pt_s}, we find 
\beq
 \frac{1}{2} ( \del_0 \del_0' + \del_1 \del_1') w_s(x; x') = \frac{1}{2 \pi} \sum_{n = 1}^\infty n e^{i n (t+i \eps)} = - \frac{1}{2 \pi (t+i \eps)^2} - \frac{1}{24 \pi} + \order(t).
\eeq
For a minimally coupled scalar field with a variable mass $m^2(x)$ in two dimensional space-time, the Hadamard parametrix is given by (see, e.g., \cite{DecaniniFolacci08})
\beq
 h(x; x') = - \frac{1}{4 \pi} \left( 1 + \frac{1}{2} m^2(x) \rho(x, x') + \order((x-x')^3) \right) \log \frac{\rho_\eps(x,x')}{\Lambda^2},
\eeq
where $\rho$ is the \emph{Synge world function}, i.e., $\frac{1}{2}$ times the squared (signed) geodesic distance of $x$ and $x'$, \cf \cite{Poisson03}, and $\Lambda$ is a length scale (the ``renormalization scale''). For the local covariance, it is crucial that $\Lambda$ is fixed and does not depend on any geometric data \cite{HollandsWaldWick}. Inside of the logarithm, the world function is equipped with an $i \eps$ prescription as follows:
\beq
 \rho_\eps(x, x') = \rho(x, x') + i \eps (\tau - \tau').
\eeq

For the scalar part, the mass term is absent. We thus find, 
for the coinciding point limit from the time direction,\footnote{Here and in the following, $\order(t)$ also includes terms of the form $t \log t$.} 
\beq
 \frac{1}{2} ( \del_0 \del_0' + \del_1 \del_1') h_s = - \frac{1}{2 \pi (t+i \eps)^2} + \order(t).
\eeq
For the scalar contribution to the energy, we thus obtain
\beq
\label{eq:H0_s}
 \VEV{H^0_s} = \int_0^{2\pi} \VEV{H^0_s(\sigma)} \ud \sigma = - \frac{1}{12}.
\eeq

For the planar contribution to the energy density, we compute
\begin{multline}
\label{eq:H0_p_w}
 \frac{1}{2} \Tr \left[ \left( \del_0 \del_0' + \del_1 \del_1' + C + B \del_1' \right) w_p(x; x') \right] \\
 = \frac{1}{2 \pi} \left[ \sum_{n = 1}^\infty n e^{i n (t+i \eps)} + 2 \sum_{n = 2}^\infty n e^{i n (t+i \eps)} \right] = - \frac{3}{2 \pi (t+i \eps)^2} - \frac{3}{24 \pi} - \frac{1}{\pi} + \order(t).
\end{multline}
In order to obtain the planar parametrix, it is convenient to absorb the first order derivatives in the equation of motion \eqref{eq:eom_p} by introducing suitable covariant derivatives, analogously to the procedure followed in \cite{BGP07} for the construction of retarded and advanced Green's functions. It is straightforward to check that with 
\begin{align}
 A_0 & = \sqrt{2} \begin{pmatrix} 0 & 1 & 0 \\ -1 & 0 & 0 \\ 0 & 0 & 0 \end{pmatrix}, &
 A_1 & = \sqrt{2} \begin{pmatrix} 0 & 0 & 1 \\ 0 & 0 & 0 \\ -1 & 0 & 0 \end{pmatrix}, &
 M^2 & = \frac{2}{R^2} \begin{pmatrix} 0 & 0 & 0 \\ 0 & -1 & 0 \\ 0 & 0 & 1 \end{pmatrix},
\end{align}
and the covariant derivative
\beq
 \nabla_\mu = \del_\mu + A_\mu,
\eeq
the planar part of the free action can be written as
\beq
\label{eq:ActionGeometric}
 \sS^0_p = - \frac{1}{2} \int \left( \bar g^{\mu \nu} \nabla_\mu f \nabla_\nu f + f^T M^2 f \right) \sqrt{- \bar g} \ud^2 x.
\eeq
We thus can interpret $M^2$ as a ``mass term''.
Denoting by $\1(x, x')$ the parallel transport from $x'$ to $x$ \wrt the covariant derivative, the parametrix for the planar sector is then given by
\begin{equation}
 h_p(x, x') = - \frac{1}{4 \pi} \left[ \1(x, x') - \frac{1}{2} \begin{pmatrix} 0 & 0 & 0 \\ 0 & 1 & 0 \\ 0 & 0 & - 1 \end{pmatrix} (x-x')^2 + \order( (x-x')^3 ) \right]  \log \frac{R^2 (x-x')^2 + i \eps ( \tau - \tau' )^2}{4 \Lambda^2},
\end{equation}
with $(x-x')^2$ the usual Minkowski square. The parallel transport can be Taylor expanded around coinciding points, i.e., in $\Delta x = x - x'$ as
\beq
\label{eq:ParallelTransportTaylor}
 \1(x, x') = \1 - A_\mu \Delta x^\mu + \frac{1}{2} \left( A_\nu A_\mu + \del_\mu A_\nu \right) \Delta x^\mu \Delta x^\nu + \order(\Delta x^3).
\eeq
Using these results, one computes
\begin{equation}
 \frac{1}{2} \Tr \left[ \left( \del_0 \del_0' + \del_1 \del_1' + C + B \del_1' \right) h_p(x; x') \right]
 = - \frac{3}{2 \pi (t+i \eps)^2} - \frac{1}{\pi} + \order(t).
\end{equation}
To obtain this result, one notes that the last two terms on the \lhs are irrelevant, due to the trace and the direction from which the coinciding point limit is taken. The first two terms on the \lhs yield
\beq
 - \frac{1}{2 \pi} \Tr \left[ \frac{\1 + t A_0 + \frac{1}{2} (A_0 A_0 + \del_0 A_0) t^2}{t^2} + \frac{-A_0 - (A_0 A_0 + \del_0 A_0) t}{t} \right],
\eeq
with the first term stemming from the action of all derivatives on the $\log$, whereas the second term stems from the action of one time derivative on $\1$ and one on the $\log$.
For the planar contribution to the energy, we thus finally obtain
\beq
 \VEV{H^0_p} = - \frac{3}{12}.
\eeq
Adding this to the scalar contribution \eqref{eq:H0_s} and using \eqref{eq:InterceptDetermination}, we obtain the intercept $a=\frac{1}{6}$, consistent with \eqref{eq:a_HS} for $D=6$. As shown in Section~\ref{sec:D}, this consistency persists to all dimensions and general ellipticity.


We note that there is no logarithmic divergence in the coinciding point limit of the world-sheet energy density, and thus no renormalization ambiguity associated to the choice of the scale $\Lambda$. The reason is that the only renormalization ambiguity at the order considered here is a supplementary Einstein-Hilbert term, which however vanishes on our flat background (it is generically irrelevant for closed strings due to the Gau\ss-Bonnet theorem), \cf also the discussion at the end of the next section.


\section{Generalization to the elliptic case}
\label{sec:Elliptic}

To treat the case of an elliptic string, we use, instead of \eqref{eq:X}, the classical solution
\begin{equation}
 \bar X(\tau, \sigma) = \begin{pmatrix} R \tau \\ R_{1,2} \sin \tau \cos \sigma \\ - R_{1,2} \cos \tau \cos \sigma \\ R_{3,4} \cos \tau \sin \sigma \\ R_{3,4} \sin \tau \sin \sigma \\ 0 \end{pmatrix},
\end{equation}
where $\sigma \in [0, 2 \pi)$ and $R$ is given by \eqref{eq:Def_R}.
We also introduce the angle $\theta \in [0, \frac{\pi}{2}]$ parameterizing $R_{1,2}$ and $R_{3,4}$ by
\begin{align}
 R_{1,2} & = R \cos \theta, &
 R_{3,4} & = R \sin \theta. \label{eq:Def_theta}
\end{align}
Two limiting cases are of particular interest: For $\theta \to \frac{\pi}{4}$, one recovers the circular case, whereas for $\theta \to 0$, one finds two straight open strings attached at the endpoints. 

The induced metric on the world-sheet, in the coordinates introduced above, is
\beq
\label{eq:metric_elliptic}
 \bar g_{\mu \nu} = \frac{R^2}{2} \left( 1 - \cos 2 \theta \cos 2 \sigma \right) \eta_{\mu \nu}.
\eeq
As this is conformal to the Minkowski metric, the equation of motion \eqref{eq:eomX} can be easily checked. 
Energy and angular momentum of the above solution were given in \eqref{eq:E_class}.
The scalar curvature is given by
\beq
\label{eq:scalarCurvature}
 \cR = \frac{8}{R^2} \frac{\cos^2 2 \theta - \cos 2 \theta \cos 2 \sigma}{(1- \cos 2 \theta \cos 2 \sigma)^3}.
\eeq
Hence, unless $\theta = \frac{\pi}{4}$, i.e., for $R_{1,2} = R_{3,4}$, the world-sheet has intrinsic curvature.

Perturbations of this classical solution are described again in the form \eqref{eq:X_phi}, \eqref{eq:phi_Parametrization}, but for the vectors $v_a$, $v_b$, $v_c$, orthonormal and normal to the world-sheet, we now choose
\begin{align}
 v_a & = \frac{1}{\cM \cN} \begin{pmatrix} \cN \\ 2 \cos \theta \cos \tau \cos \sigma \\ 2 \cos \theta \sin \tau \cos \sigma \\ - 2 \sin \theta \sin \tau \sin \sigma \\ 2 \sin \theta \cos \tau \sin \sigma \\ 0 \end{pmatrix}, &
 v_b & = \frac{\sqrt{2}}{\cM} \begin{pmatrix} 0 \\ \sin \theta \sin \tau \cos \sigma \\ - \sin \theta \cos \tau \cos \sigma \\ \cos \theta \cos \tau \sin \sigma \\ \cos \theta \sin \tau \sin \sigma \\ 0 \end{pmatrix}, &
 v_c & = \frac{\sqrt{2}}{\cN} \begin{pmatrix} 0 \\ \sin \theta \cos \tau \sin \sigma \\ \sin \theta \sin \tau \sin \sigma \\ \cos \theta \sin \tau \cos \sigma \\ - \cos \theta \cos \tau \cos \sigma \\ 0 \end{pmatrix},
\end{align}
where
\begin{align}
 \cM & = \sqrt{1 - \cos 2 \theta \cos 2 \sigma}, & \cN & = \sqrt{1 + \cos 2 \theta \cos 2 \sigma}.
\end{align}
In the limit of the circular string, these coefficients tend to $1$ and the above vectors coincide with the vectors used in Section~\ref{sec:ClassicalSolutions}.

The free part of the action still has the form \eqref{eq:S0}, but now with $A$, $B$, $C$ replaced by
\begin{subequations}
\begin{align}
 A & = \frac{2}{\cM^2 \cN} \begin{pmatrix} 0 & \sqrt{2} \sin 2 \theta & 0 \\ -\sqrt{2} \sin 2 \theta & 0 & \cM \cos 2 \theta \sin 2 \sigma \\ 0 & - \cM \cos 2 \theta \sin 2 \sigma & 0 \end{pmatrix}, \\
 B & = \frac{\sqrt{8} \sin 2 \theta}{\cM \cN^2} \begin{pmatrix} 0 & 0 & - 1 \\ 0 & 0 & 0 \\ 1 & 0 & 0 \end{pmatrix}, \\
 C & = \frac{1}{\cM^4 \cN^4} \begin{pmatrix} - \cB/4 & 0 & 3 \cM \cN^2 \sin 4 \theta \sin 2 \sigma / \sqrt{2} \\ 0 & -\cN^4 ( \cM^4 + 3 \sin^2 2 \theta ) & 0 \\ 3 \cM \cN^2 \sin 4 \theta \sin 2 \sigma / \sqrt{2} & 0 & - \cM^4 \cN^4 + \cM^2 \sin^2 2 \theta ( 4 \cN^2 + \cM^2 ) \end{pmatrix},
\end{align}
\end{subequations}
where we denoted
\beq
\BB=\cos 2\theta\cos 2\sigma-5\cos 6\theta\cos 2\sigma+8\cos^2 2\theta(\cos 4\sigma-1)+4\cos^32\theta\cos 6\sigma.
\eeq
The equation of motion \eqref{eq:eom_s} for the scalar part is not modified, whereas the equation of motion \eqref{eq:eom_p} for the planar part is replaced by
\beq
 0 = - \ddot f + f'' - A \dot f - B f' - (C + B'/2) f,
\eeq
the modification being due to the fact that $B$ is no longer constant. The symplectic form of the planar part is still of the form \eqref{eq:SymplecticForm}, with the modified $A$, of course. The free Hamiltonian is still given by \eqref{eq:H_0}, again with the modified $B$, $C$.

The symplectically normalized mode solutions for the scalar excitations are still given by \eqref{eq:f_s_n}, \eqref{eq:f_q_f_p}. Those for the planar sector are tedious to derive. The results are stated in the appendix.
For our purposes, the important point about these mode solutions is not their explicit form, but that they have exactly the same frequencies as in the circular case. Canonical quantization proceeds in complete analogy to the circular case. Renormalization, however, is affected by the curvature of the elliptic string. We aim to perform it in a locally covariant way, i.e., with respect to the local geometric data. Hence, we rewrite the free action in a geometric form as in \eqref{eq:ActionGeometric}, now with
\begin{subequations}
\label{subeq:A_M2_elliptic}
\begin{align}
\label{eq:A_elliptic}
 A_0 & = \sqrt 2 \begin{pmatrix}  0 & \frac{\sin 2\theta}{\MM^2\NN} & 0 \\ -\frac{\sin 2\theta}{\MM^2\NN} & 0 & \frac{\cos 2\theta \sin 2\sigma}{\sqrt{2} \MM\NN} \\ 0 & -\frac{\cos 2\theta \sin 2\sigma}{\sqrt{2} \MM\NN}  & 0 \end{pmatrix}, &
A_1 & = \sqrt 2 \begin{pmatrix} 0 & 0 & \frac{\sin 2\theta}{\MM\NN^2}\\ 0 & 0 & 0 \\ -\frac{\sin 2\theta}{\MM\NN^2}& 0 & 0 \end{pmatrix}, \\
\label{eq:M2_elliptic}
M^2 & = \frac{2}{R^2\MM^2} \begin{pmatrix} \frac{4\cos^2{2\theta}\sin^2{2\sigma}}{\MM^4\NN^2} & 0 & \frac{\sqrt{2}\sin{4\theta}\sin{2\sigma}}{\MM^3\NN^2} \\ 0 & -\frac{2\sin^2{2\theta}}{\MM^4} & 0 \\ \frac{\sqrt{2}\sin{4\theta}\sin{2\sigma}}{\MM^3\NN^2} & 0 & \frac{2\sin^2{2\theta}}{\MM^2\NN^2} \end{pmatrix}.
\end{align}
\end{subequations}
For the scalar part, one proceeds analogously, with the derivative the usual one, and a vanishing ``mass'' $M^2$.
Furthermore, as the metric is no longer flat, but given by \eqref{eq:metric_elliptic}, the expression for the world function becomes more complicated. It can be Taylor expanded in coordinates around coinciding points as \cite{OttewillWardell}
\begin{subequations}
\begin{align}
 \rho(x,x') & = \tfrac{1}{2} \bar g_{\mu \nu}(x) \Delta x^\mu \Delta x^\nu + A_{\mu \nu \lambda}(x) \Delta x^\mu \Delta x^\nu \Delta x^\lambda + B_{\mu \nu \lambda \rho}(x) \Delta x^\mu \Delta x^\nu \Delta x^\lambda \Delta x^\rho, \\
 A_{\mu \nu \lambda} & = - \tfrac{1}{4} \del_{(\mu} \bar g_{\nu \lambda)}, \\
 B_{\mu \nu \lambda \rho} & = \tfrac{1}{12} \del_{(\mu} \del_{\nu} \bar g_{\lambda \rho)} - \tfrac{1}{24} g^{\sigma \tau} \left( \tfrac{1}{4} \del_\sigma \bar g_{(\mu \nu} \del_{|\tau|} \bar g_{\lambda \rho)} - \del_{\sigma} \bar g_{(\mu \nu} \del_{\lambda} \bar g_{\rho) \tau} + \del_{(\mu} \bar g_{\nu | \sigma|} \del_{\lambda} \bar g_{\rho) \tau} \right),
\end{align}
\end{subequations}
where $\Delta x = x - x'$. One thus finds, for a metric of the form $\bar g_{\mu \nu} = f(\sigma) \eta_{\mu \nu}$,
\begin{multline}
\label{eq:rhoTaylor}
 \rho = \tfrac{1}{2} f(\sigma) \left( - \Delta \tau^2 + \Delta \sigma^2 \right) + \tfrac{1}{4} f'(\sigma) \Delta \tau^2 \Delta \sigma - \tfrac{1}{96} f(\sigma)^{-1} f'(\sigma)^2 \Delta \tau^4 \\ + \left( \tfrac{1}{48} f(\sigma)^{-1} f'(\sigma)^2 - \tfrac{1}{12} f''(\sigma) \right) \Delta \tau^2 \Delta \sigma^2 + \order(\Delta x^5, \Delta \sigma^3),
\end{multline}
and hence,
for the coinciding point limit from the time direction, 
\beq
 \frac{1}{2} ( \del_0 \del_0' + \del_1 \del_1') h_s = - \frac{1}{2 \pi (t+i \eps)^2} + \frac{1}{32 \pi} \frac{f'^2}{f^2} - \frac{1}{48 \pi} \frac{f''}{f} + \order(t).
\eeq
For the metric \eqref{eq:metric}, this yields
\beq
 \frac{1}{2} ( \del_0 \del_0' + \del_1 \del_1') h_s = - \frac{1}{2 \pi (t+i \eps)^2} + \frac{\cos^2 2 \theta \sin^2 2 \sigma}{8 \pi (1-\cos 2 \theta \cos 2 \sigma)^2} - \frac{\cos 2 \theta \cos 2 \sigma}{12 \pi (1-\cos 2 \theta \cos 2 \sigma)} + \order(t).
\eeq
For the scalar contribution to the energy density, we thus obtain
\beq
 \VEV{H^0_s(\sigma)} = - \frac{1}{24 \pi} - \frac{\cos^2 2 \theta \sin^2 2 \sigma}{8 \pi (1-\cos 2 \theta \cos 2 \sigma)^2} + \frac{\cos 2 \theta \cos 2 \sigma}{12 \pi (1-\cos 2 \theta \cos 2 \sigma)}.
\eeq
In the degenerate limit $\theta \to 0$, this gives
\beq
\label{eq:thetaTo0}
 \lim_{\theta \to 0} \VEV{H^0_s(\sigma)} = - \frac{1}{12 \pi \sin^2 \sigma},
\eeq
in accordance with the result obtained for the open string \cite{RotatingStringEnergy}. Integration over $\sigma$ yields
\beq
\label{eq:H0_h_s_elliptic}
 \VEV{H^0_s} = - \frac{1}{12 \sin 2 \theta} = - \frac{1}{24} \left( \tan \theta + \cot \theta \right).
\eeq

For the planar part, the computation of the expectation value of the Hamiltonian density is challenging. However, we are only interested in the total Hamiltonian. Noting that it can still be expressed in terms of creation and annihilation operators as in \eqref{eq:H_0_ladder}, the computation of the expectation value, point-split in time direction, proceeds, up to the factor $2 \pi$, as in the second line of \eqref{eq:H0_p_w}. For the contribution from the two-point function, we thus obtain
\beq
\label{eq:H0_w_p_elliptic}
 -\frac{3}{(t+\ii\epsilon)^2}-\frac{3}{12} - 2.
\eeq
The parametrix for the planar part is 
\beq
 h_p(x; x') = - \frac{1}{4 \pi} \left( \1(x,x') + \frac{1}{2} M^2(x) \rho(x, x') + \order((x-x')^3) \right) \log \frac{\rho_\eps(x,x')}{\Lambda^2},
\eeq
with $M^2$ and the parallel transport now defined by \eqref{subeq:A_M2_elliptic}. Using \eqref{eq:ParallelTransportTaylor} and \eqref{eq:rhoTaylor}, one finds
\begin{multline}
\label{eq:H0_h_p_sigma_elliptic}
 \frac{1}{2} \Tr \left[ \left( \pd_0 \pd'_0 + \pd_1 \pd'_1 + C + B \pd'_1 \right) h_p(x;x')\right] \\
 =-\frac{3}{2\pi(t+\ii\epsilon)^2}-\frac{\sin^2{2\theta}(17-5\cos{4\sigma})+80(\cos^4\sigma\sin^4\theta+\cos^4\theta\sin^4\sigma)}{32\pi\MM^4}\\
-\frac{\cos{2\theta}(\cos{2\theta}-\cos{2\sigma})}{2\pi\MM^4}\log{\left[-\frac{R^2(t+\ii\eps)^2\MM^2}{4\Lambda^2}\right]}+\order(t).
\end{multline}
Integrating this over $\sigma$, we obtain
\begin{equation}
\label{eq:H0_h_p_elliptic}
-\frac{3}{(t+\ii\epsilon)^2}-\frac{1}{4}-\frac{7}{8}(\tan\theta+\cot\theta).
\end{equation}
Subtracting now \eqref{eq:H0_h_p_elliptic} from \eqref{eq:H0_w_p_elliptic}, we obtain the expression for the planar contribution to the free Hamiltonian expectation value
\begin{equation}
\VEV{H^0_p}=-2+\frac{7}{8}(\tan\theta+\cot\theta).
\end{equation}

As a consistency check, we note that the coefficient of the logarithm in the last term in \eqref{eq:H0_h_p_sigma_elliptic} can be also expressed, using \eqref{eq:scalarCurvature}, as
\beq
-\frac{\cR \sqrt{\bar g}}{8\pi},
\eeq
meaning that a change in the renormalization scale $\Lambda$ corresponds to finite renormalization of the Einstein-Hilbert term. By the analysis of \cite{HollandsWaldWick} this is the only renormalization ambiguity for the world sheet energy density $H^0(\sigma)$ in a locally covariant renormalization scheme.
For the closed string, the contribution of this term to the Hamiltonian $H^0$ vanishes, by the Gau\ss-Bonnet theorem, so that the obtained result is unambiguous.

\section{The generalization to arbitrary dimensions}
\label{sec:D}

For the case of general dimension $D$, one has to consider $D-5$ scalar fields, i.e., instead of $v_s$ defined in \eqref{eq:Def_v}, we consider $v_{s,i} = e_{4+i}$, for $i = 1, \dots, D-5$ and $e_j$ the canonical basis vectors of $(\R^D, \eta)$. Instead of the single coefficient $f_s$, we also have $f_{s, i}$. With these, the free part of the action \eqref{eq:S0} is modified to
\begin{equation}
\label{eq:S0_D}
 \sS^0 = \frac{1}{2} \int \left( {\dot f}^2 - {f'}^2 - f^T A \dot f - f^T B f' - f^T C f + \sum_{i = 1}^{D-5} \left[ {\dot f_{s, i}}^2 - {f'_{s, i}}^2 \right] \right) \ud \sigma \ud \tau.
\end{equation}
We see that at this order, the perturbations in the directions orthogonal to the rotation subspace spanned by $e_0 - e_4$ completely decouple, i.e., we are dealing with $D-5$ independent copies of the scalar field previously considered. It follows that vacuum expectation value of the free Hamiltonian in $D$ target space dimensions is
\beq
 \VEV{H^0}=(D-5) \VEV{H^0_s}+\VEV{H^0_p}=-2-\frac{D-26}{24}(\tan\theta+\cot\theta).
\eeq
Using \eqref{eq:InterceptDetermination}, we obtain the Regge intercept
\begin{equation}
a=1+\frac{D-26}{48}(\tan\theta+\cot\theta).
\end{equation}
Using that by \eqref{eq:Def_theta} and \eqref{eq:E_class} we have $\cot \theta = \sqrt{J_{12} / J_{34}}$, this result can be seen to coincide with the result \eqref{eq:a_HS} obtained in \cite{HellermanSwanson}.

Expanding the action beyond second order in the perturbation, the different scalar modes $f_{s, i}$ will of course couple among each other and with the planar modes, so the simple linear scaling with the dimension is expected not to hold beyond the semi-classical approximation.

Let us finally comment on the origin of the divergence for $\theta \to 0, \frac{\pi}{2}$, i.e., $J_{3/4} \to 0$ or $J_{1/2} \to 0$. In this limit, the closed string is flattened to two straight open strings joined at the folds $\sigma = 0, \pi$. In particular, the scalar curvature \eqref{eq:scalarCurvature} diverges at these folds. This leads to a non-integrable divergence of the local energy density at the folds, as seen in \eqref{eq:thetaTo0}. In the case of the open string \cite{RotatingStringEnergy}, the same non-integrable divergence occurred, but it could be removed using geodesic curvature boundary counterterms. This procedure can not be applied in the present case, as the closed string does not possess boundaries.\footnote{For $\theta = 0$ one could of course define the energy to be twice that of a single open string, but that does not remedy the divergence as $\theta \to 0$.}

\section{The spectrum of physical excitations}
\label{sec:ExcitationSpectrum}

Let us recall some ingredients of the covariant quantization of the closed Nambu-Goto string. To simplify notation, we set $\pi \gamma = 1$. We have two sets $A_m^{\pm \mu}$ of oscillators, for right- and left-moving excitations, fulfilling
\beq
 [A^{\alpha \mu}_m, A^{\beta \nu}_n] = m \delta_{m + n} \delta^{\alpha \beta} \eta^{\mu \nu}.
\eeq
One identifies
\beq
 p^\mu = \frac{1}{2} A^{\pm \mu}_0 
\eeq
as the center of mass momentum.
There are also two sets $L^\pm_m$ of Virasoro generators. Furthermore, there is the angular momentum operator
\beq
 J^{\mu \nu} = - i \sum_{\alpha \in \pm} \sum_{n \geq 1} \frac{1}{n} \left( A^{\alpha \mu}_{-n} A^{\alpha \nu}_n - \mu \leftrightarrow \nu \right).
\eeq
One has the commutation relations
\begin{align}
[L^\alpha_m,\zeta \cdot A^\beta_{-k}] & = k \delta^{\alpha \beta} \zeta \cdot A^\alpha_{m-k}, &
[J^{12},\zeta \cdot A^\alpha_{-k}] & = \zeta^{12} \cdot A^\alpha_{-k}, &
[J^{34},\zeta \cdot A^\alpha_{-k}] & = \zeta^{34} \cdot A^\alpha _{-k},
\end{align}
with $\zeta = (\zeta^0, \dots, \zeta^{D-1})$ a target space vector and
\begin{equation}
\zeta^{12}=(0,-\ii\zeta^2,\ii\zeta^1,0,\ldots,0),\qquad\zeta^{34}=(0,0,0,-\ii\zeta^4,\ii\zeta^3,0,\ldots,0).
\end{equation}
We also recall that physical states $\Phi$ have to fulfill the conditions
\beq
\label{eq:Physicality}
 (L^\pm_m - \delta_m a ) \ket{\Phi} = 0
\eeq
for all $m \geq 0$. In particular, we recall that
\beq
 L^\pm_0 = \frac{1}{8} p^2 + \sum_{n \geq 1} A^\pm_{-n} \cdot A^\pm_n.
\eeq

The physical state of minimum energy for given angular momentum $\ell_{1,2}$ in the 1-2 plane and $\ell_{3,4}$ in the 3-4 plane, with $\ell_{1,2}$, $\ell_{3,4}$ positive even numbers, is given by
\begin{equation}
\label{eq:GroundStateCovariant}
\ket{\ell_{1,2}, \ell_{3,4}} = (\xi \cdot A^+_{-1})^{\ell_{1,2}/2} (\tilde \xi \cdot A^+_{-1})^{\ell_{3,4}/2} (\xi \cdot A^-_{-1})^{\ell_{1,2}/2} (\tilde \xi \cdot A^-_{-1})^{\ell_{3,4}/2} \ket{0; \sqrt{2(2 \ell_{1,2} + 2 \ell_{3,4} - 4 a)}},
\end{equation}
with $\ket{0;M}$ denoting a ground state with vanishing spatial momentum and rest mass $M$ and
\begin{equation}
\xi=\frac{1}{\sqrt{2}}(0,1,\ii,0,\ldots,0),\qquad\tilde\xi=\frac{1}{\sqrt{2}}(0,0,0,1,\ii,0,\ldots,0).
\end{equation}
Note that we have to evenly excite left- and right-movers in order to fulfill the physicality condition \eqref{eq:Physicality} for $m = 0$. We thus recover the Regge trajectory \eqref{eq:ReggeTrajectory}. We introduce the notation
\beq
 M_\ell^2 = 2 \left( 2 \ell_{1,2} + 2 \ell_{3,4} - 4 a \right).
\eeq

Let us discuss what the analog of a scalar excitation of the state \eqref{eq:GroundStateCovariant} would look like. Naively, we would choose a unit vector $\theta$ in the subspace orthogonal to the one spanned by $X^0$ --- $X^4$, and act with the creation operator $\theta \cdot A_{-1}$ (and adjust the momentum, i.e., the mass accordingly). However, in order to fulfill both equations \eqref{eq:Physicality} for $m=0$ with the same $a$, we also have to excite a right-moving mode. It follows that minimal excitations of \eqref{eq:GroundStateCovariant} in the directions perpendicular to $X^0$ --- $X^4$ are given by
\beq
\ket{\ell_{1,2}, \ell_{3,4}, \theta, \tilde \theta} = (\theta \cdot A^+_{-1}) (\tilde \theta \cdot A^-_{-1} ) (\xi \cdot A^+_{-1})^{\ell_{1,2}/2} (\tilde \xi \cdot A^+_{-1})^{\ell_{3,4}/2} (\xi \cdot A^-_{-1})^{\ell_{1,2}/2} (\tilde \xi \cdot A^-_{-1})^{\ell_{3,4}/2} \ket{0; \sqrt{M_\ell^2 + 8}}.
\eeq
Here we adjusted the rest mass in order to fulfill the physicality condition \eqref{eq:Physicality}. We see that in the covariant scheme, the minimal scalar excitation leads to a shift
\beq
 \Delta M^2 = 8
\eeq
of the rest mass squared.

Let us now consider the situation in the semiclassically quantized string. As is obvious from \eqref{eq:H_0_ladder}, the minimal scalar excitation, $\hat a^{+ *}_{s, 1}$ or $\hat a^{- *}_{s, 1}$, raises the expectation value of $H^0$ by $1$. According to \eqref{eq:E2}, and taking $\pi \gamma =1$ into account, this leads to a shift
\beq
 \Delta M^2 = 4
\eeq
of the rest mass squared. We thus see that, at the linearized level considered here, the semiclassically quantized string exhibits physical excitations\footnote{We recall that at the order considered here, we could set all unphysical (auxiliary and pure gauge) fields to zero. The remaining ones are physical (commute with the free BRST charge) and likewise are the corresponding excitations.} that are not present in the covariantly quantized string. The point is that in the semiclassical scheme (at the linearized level) no condition on the equal excitation of left- and right-movers is present. Presently, it is unclear whether this is remedied at higher order in perturbation theory or not (and thus represents a non-perturbative effect).

\section{Conclusion}
\label{sec:Conclusion}

We computed semi-classical corrections to the energy of rotating closed Nambu-Goto strings. Special care was taken to renormalize in a local and covariant way, using techniques developed in the context of QFT on curved space-times. This calculation represents one of the few analytically tractable non-trivial applications of these techniques (needed for example for the calculation of backreaction effects in cosmological \cite{HackBook} or black hole spacetimes \cite{ParkerToms}). Our results agree with those obtained \cite{HellermanSwanson} via the Polchinski-Strominger action, which provides a mutual consistency check.

The results obtained imply a discrepancy with the covariant quantization scheme for $D<26$, where the intercept $a$ is only constrained by $a \leq 1$, but independent of the ellipticity. In contrast, in \cite{HellermanSwanson} and the present work, the intercept $a$ is computed, but found to depend on the ellipticity for $D \neq 26$. We also showed that in our linearized approximation there are physical excitations not present in the covariant quantization scheme, even for $D=26$. It would be interesting to learn whether this is remedied at higher order in perturbation theory, similar to linearized Yang-Mills theory containing more observables than the full non-linear theory, or whether this constitutes a non-perturbative effect.

\appendix 
\section{The planar modes in the elliptic case}

The (not symplectically orthonormalized) positive frequency planar mode solutions in the elliptic case are given by
\begin{subequations}
\begin{align}
f^\pm_{1,n} & =\frac{1}{\MM\NN} \bigg[\sqrt{2}\left(\pm \sqrt{2}\ii,\mp n\NN\sin{2\theta},n\MM\sin{2\theta} \right)e^{\pm\ii n\sigma} +\left(\pm \ii (n+1)\cos{2\theta},0,0\right) e^{\pm\ii (n-2)\sigma} \nn \\
\label{eq:f_1_n_elliptic}
&\qquad \qquad +\left(\mp \ii (n-1)\cos{2\theta},0,0\right)e^{\pm\ii (n+2)\sigma} \bigg]e^{-\ii n\tau}, \\
f^\pm_{2,n} & =\frac{1}{\MM\NN} \bigg[2\cos{2\theta}\left(\pm 2\ii,\mp n\NN,n\MM \right)e^{\pm\ii n\sigma} +\left(\mp2 \ii (n-1)\cos^2{2\theta},0,0\right)e^{\pm\ii (n+2)\sigma} \nn \\
\label{eq:f_2_n_elliptic}
& \qquad \qquad + \left(\pm \ii \cA_-,\pm 2n\NN(1-\sqrt{2}\sin{2\theta}), 2n\MM(1-\sqrt{2}\sin{2\theta})\right) e^{\pm\ii (n-2)\sigma} \bigg]e^{-\ii n\tau}, \\
f^\pm_{3,n} & =\frac{1}{\MM\NN} \bigg[2\cos{2\theta}\left(\pm 2\ii,\mp n\NN,n\MM \right)e^{\pm\ii n\sigma} +\left(\pm 2 \ii (n+1)\cos^2{2\theta},0,0\right)e^{\pm\ii (n-2)\sigma} \nn \\
\label{eq:f_3_n_elliptic}
& \qquad \qquad + \left(\pm \ii \cA_+,\pm 2n\NN(1-\sqrt{2}\sin{2\theta}), 2n\MM(1-\sqrt{2}\sin{2\theta})\right)e^{\pm\ii (n+2)\sigma}  \bigg]e^{-\ii n\tau},
\end{align}
\end{subequations}
where $n \geq 2$ and
\beq
\cA_\mp = \left(1\pm3n\mp(n\mp1)\cos{4\theta}\mp\sqrt{8}n\sin{2\theta}\right).
\eeq
The mode solution \eqref{eq:f_3_n_elliptic} can also be extended to $n=1$. The zero energy mode $f^\pm_{3,0}$ is now given by
\begin{multline}
 f^\pm_{3,0} = 4 \cos{2\theta}\left(\pm\frac{\sqrt{8}\ii}{\MM\NN},\mp\frac{1}{\MM},\frac{1}{\NN}\right) \\
+\left(\pm\sqrt{2}\ii\frac{3+\cos{4\theta}\mp 2\sin{2\theta}}{\MM\NN},\pm 2\frac{(\cos\theta\mp\sin\theta)^2}{\MM},2\frac{(\cos\theta\mp\sin\theta)^2}{\NN}\right)e^{-2\ii\sigma}\\
+\left(\pm\sqrt{2}\ii\frac{3+\cos{4\theta}\pm 2\sin{2\theta}}{\MM\NN},\pm2\frac{(\cos\theta\pm\sin\theta)^2}{\MM},2\frac{(\cos\theta\pm\sin\theta)^2}{\NN}\right)e^{2\ii\sigma}.
\end{multline}
The modes $f^\pm_q$, which we interpreted as position operators in the planes of rotation, are now given by
\beq
 f_q^\pm= \frac{1}{\cM \cN} \left[ \left(\pm \sqrt{2}\ii, - \cN, \cM \right) (\cos\theta \pm \sin\theta) e^{\ii\sigma} + \left(\pm \sqrt{2}\ii, \cN, \cM \right)(\cos\theta\mp\sin\theta)e^{-\ii\sigma} \right] e^{-\ii\tau}.
\eeq
There is also a corresponding linearly growing mode $f_p^\pm$, however, for its homogeneous part, i.e., the part not linearly growing in time, we were only able to derive an ODE, and not to explicitly solve it. Likewise, for the generalization of the mode $f_\theta$ describing rotations generated by $L_{1,2} + L_{3,4}$, we find
\beq
 f_\theta=\left(\frac{2\left(\cos^4 \theta \cos^2\sigma+\sin^4 \theta\sin^2\sigma\right)}{\MM\NN}, 0, \frac{ \sin{4\theta}\sin {2\sigma}}{4\sqrt{2} \NN}\right),
\eeq
again with an unknown homogeneous part of the corresponding linearly growing mode $f_\lambda$.


\end{document}